\documentclass[aps,prd,showpacs]{revtex4}
\usepackage{graphicx}
\usepackage{epsfig}
\usepackage{psfig}
\usepackage{revsymb}
\begin{document}
\renewcommand{\thesection}{\arabic{section}}
\renewcommand{\thesubsection}{\arabic{subsection}}
\title{Work Function of Strongly Magnetized Neutron Star Crustal Matter and the Associated
Magneto-Sphere}
\author{Arpita Ghosh and Somenath Chakrabarty$^\dagger$}
\affiliation{
Department of Physics, Visva-Bharati, Santiniketan 731 235, 
West Bengal, India\\ 
$^\ddagger$E-mail:somenath.chakrabarty@visva-bharati.ac.in}
\begin{abstract}
Following an extremely interesting idea \cite{R1}, published long ago, the work function at the outer crust region of a strongly magnetized neutron star is obtained using relativistic version of Thomas-Fermi type
model. In the present scenario, the work function becomes anisotropic; the longitudinal part is an increasing function of magnetic field strength, whereas the transverse part diverges. An approximate estimate of the electron density in the magnetosphere due to field emission and photo emission current, from the
polar cap region are obtained.
\end{abstract}
\maketitle
The study of the formation of plasma in a pulsar magnetosphere is a quite old but still an unresolved
astrophysical issue. In the formation of magnetosphere plasma, it is generally assumed that there must
be an initial high energy electron flux from the magnetized neutron star. At the
poles of a neutron star the emitted charged particles flow only along the magnetic field lines. Further a rotating magnetized neutron star generates extremely high electro-static potential difference near the poles.
This potential difference is the driving force and plays the major role in the extraction of electrons
as field emission or what is also called cold emission, from the crustal matter of strongly magnetized
neutron stars near the poles. The flow of high energy electrons along the direction of magnetic field
lines and their penetration through the light cylinder is pictured with the current carrying
conductors. Naturally, if the conductor is broken near the pulsar surface the entire potential
difference will be developed across a thin gap, called polar gap. This is based on the assumption that
above a critical height, from the polar cap, because of high electrical conductivity of the plasma, the electric field $E_{\vert\vert}$ ($=E_0$ in this article), parallel with the magnetic field near the poles is
quenched. Further, a steady acceleration of electrons originating at the polar region of neutron
stars, travelling along the field lines, will produce magnetically convertible curvature $\gamma$-rays.
If these curvature $\gamma$-ray photons have energies $>2m_ec^2$, then pairs of $e^--e^+$ will be
produced in enormous amount with very high efficiency near the polar gap. These produced $e^--e^+$
pairs form what is known as the magnetospheric plasma \cite{R2,R3,R4,R5,R6,R7,R8}.  

The process of extracting electrons from the outer crust region of strongly magnetized neutron stars,
including the most exotic stellar objects, the magnetars, one requires a more or less exact description
of the structure of matter in that region \cite{R9,R10,R11}. From the knowledge of structural
deformation of atoms in strong magnetic field; the departure from spherical nature to a
cigar shape, allows us to assume that the atoms in the outer crust region, which are fully ionized
because of high density, may be replaced by Wigner-Seitz type cells of approximately cylindrical in
structure. We further assume that the electron gas inside the cells are strongly degenerate and are at
zero temperature. It is well known that the presence of extraordinarily large magnetic field not only
distorts the crystalline structure of dense metallic iron, also significantly modifies the electrical properties of such matter. As for example, the electrical conductivity, which is otherwise isotropic, becomes highly anisotropic in presence of strong quantizing magnetic field. 
In presence of strong magnetic field iron crystal is highly conducting in the direction parallel to
the magnetic field, whereas flow of current in the perpendicular direction is severely inhibited. The
aim of this letter is to show that the work function, which is the most important parameter associated
with the emission of electrons from the polar region of strongly magnetized neutron stars, will also show anisotropy in presence of strong magnetic field.
In this article we have shown that the work function, associated with the emission of electrons along the field lines increases with  magnetic field strength.
Whereas its transverse component diverges, irrespective of the dimension of the cylindrically deformed
atoms. The scenario is very much analogous to the charge transport mechanism in presence of strong
magnetic field.
To the best of our knowledge, the study of anisotropic nature of work function in presence
of strong quantizing magnetic field, which has relevance, specially in the case of magnetized neutron
star crustal region has not been studied before. 
It is also well known that the most important surface emission processes are
thermal emission, may be enhanced by Schottky effect, field emission, caused by strong electric field
at the poles and perhaps the other important process is the photoelectric emission induced by high energy 
cosmic photons. The last process of electron emission, if at all possible, has not been studied in this
context. 

In this letter, following \cite{R1} we shall obtain work function associated with the emission of electrons,
along and transverse to the direction of magnetic field lines, and show that the later component is
infinitely large in this model. In \cite{R1} the author have obtained work function related to
emission of electrons from cold cathodes. It was associated with some engineering problems.
We assume cylindrically deformed atoms in the crustal region with their curved surfaces parallel to
the neutron star surface. Which further means, that locally, the field lines and axes of the cylinders are 
parallel with each other. To obtain the work function for both the cases, let us consider fig.(1).
Here the magnetic field is along z-axis and r-axis is orthogonal to the direction of magnetic field lines. 
To get the longitudinal part of the work function, we
assume that an electron has come out from the cylindrically distorted atom, through one of the plane
surfaces, and is at the point $P$. The electro-static potential at $Q$ produced by this electron is given by  
\begin{equation}
\phi^{(e)}(r,z)=\frac{e}{[r^2+(z-h)^2]^{1/2}}
\end{equation}	
where $AP=h$, $PD=r$ and $CQ=z$. This equation can also be expressed in the following form 
\begin{eqnarray}
\phi^{(p)}(r,z)=e\int_{\xi=0}^\infty J_0(\xi r)&&\exp[\mp(z-h)\xi]d\xi\nonumber \\ && ~~~~{\rm{for}}~~ z> ~~{\rm{or}}~~
<h
\end{eqnarray}
For the charge distribution within the distorted atom, the Poisson's equation is given by 
$\nabla^2\phi=4\pi en_e$
where $\phi$ is the electro-static potential and $n_e$ is the electron density, given by 
\begin{equation}
n_e=\frac{eB}{2\pi^2}\sum_{\nu=0}^{\nu_{max}}(2-\delta_{\nu 0})(\mu^2-m^2-2\nu eB)^{1/2}
\end{equation}
where $\nu$ is the Landau quantum number with $\nu_{max}$ the upper limit. The first factor within the
sum takes care of singly degenerate $0$th Landau level and doubly degenerate other levels with $\nu\neq 0$
and $\mu$ is the chemical potential for the electrons, given by 
$\mu=E_F-e\phi (r)={\rm{constant}}$,
this is the so called Thomas-Fermi condition, $E_F$ is the Fermi energy for the electrons (throughout
this article we have taken $\hbar=c=1$). 
Hence we can re-write the Poisson's equation in the following form
\begin{equation}
\nabla^2\phi=\frac{2e^2 B}{\pi}\sum_{\nu=0}^{\nu{max}}(2-\delta_{\nu 0})[(\mu+e\phi)^2-m_e ^2-2\nu
eB]^{1/2}
\end{equation}
To get an analytical solution for $\phi$ in cylindrical coordinate system $(r,\theta,z)$, with
azimuthal symmetry, we put, $m_e=0$, i.e., the kinetic energy is assumed to be high enough and
$\nu_{max}=0$, which is a valid approximation, provided the magnetic field is too high, so that the electrons occupy only their $0$th Landau level. Now defining $\psi=\mu+e\phi$, the solution in the cylindrical
coordinate system is given by 
\begin{eqnarray}
\psi(r,z)=\int_{\xi=0}^\infty J_0(\xi r)a(\xi)&&\exp[(\xi^2+\lambda^2)^{1/2}z]d\xi\nonumber \\ &&
~~{\rm{with}}~~ z<0
\end{eqnarray}
Here $\lambda^2=2e^3 B/\pi$ and $a(\xi)$ is some unknown spectral function. Now following \cite{R1}, we assume that there exist a fictitious secondary field in vacuum, the work function is defined as the work done by this field in pulling out an electron from the material surface to infinity. The secondary field is expressed in coherent with $\phi^{(p)}(r,z)$ and $\phi(r,z)$, and is given by
\begin{eqnarray}
\phi^{(s)}(r,z)=e\int_{\xi =0}^\infty J_0(\xi r)f(\xi)&&\exp(-\xi z)d\xi \nonumber \\ &&~~{\rm{with}}~~ z>0
\end{eqnarray}
where $f(\xi)$ is again some unknown spectral function. 
Then following \cite{R1}, using the continuity conditions of tangential and transverse components of electric field and the displacement vector respectively, we finally get
\begin{equation}
W_f=\frac{\lambda}{3}e^2=\frac{1}{3}\left (\frac{2B}{\pi B_c^{(e)}}\right )^{1/2}m_ee^3
\end{equation}
where $B_c^{(e)}\approx 4.43\times 10^{13}$G, the critical field strength for electrons to populate
their Landau levels in the relativistic scenario. This equation gives the variation of work function with the strength of magnetic field ($\sim B^{1/2}$) associated with the emission of electrons along the direction of
magnetic field. From this expression it is also obvious that for a given magnetic field strength, if
$m_e$ is replaced by $m_p$, the proton mass or $m_I$, mass of the ions, then, since $m_I\gg m_p\gg m_e$, 
the same kind of inequalities are also valid for the work functions associated with the emissions along $z$-direction. Then obviously, it needs high temperature for thermo-ionic emission of protons or ions, high
electric field for their field emissions and high frequency incident photons are essential for the
photo-emission of these heavier components.

Let us now consider the emission of electrons in the transverse direction. As shown in fig.(1),
$p$ is the position of an electron, came out through the curved surface, then the electro-static potential at $q$ is given by 
\begin{equation}
\phi^{(p)}(r,z)=\frac{e}{[z^2+(r-r_0)^2]^{1/2}}
\end{equation}
where $ap=r_0$, $cq=r$ and $pd=z$. It can also be expressed as 
\begin{eqnarray}
\phi^{(p)}(r,z)=e\int_{\xi =0}^\infty &&J_0[\xi(r-r_0)]\exp{(\mp z\xi)}d\xi \nonumber \\ &&~~{\rm{for}}~~ z>0
~~{\rm{or}}~~ z<0
\end{eqnarray}
whereas $\phi(r,z)$ and $\phi^{(s)}(r,z)$ remain unchanged. Then as before, from the continuity
conditions on the curved surface, and using some elementary formulas for the Bessel functions, we finally get
\begin{equation}
f(\xi)=\frac{\left [F\left (1+\frac{\lambda^2}{\xi^2}\right )^{1/2}-G\right]} {\left[
1-\left(1+\frac{\lambda^2}{\xi^2}\right)^{1/2}\right]}
\end{equation}
where $F=J_1[(R-r_0)\xi]/J_1(R\xi)$ and  $G=J_0[(R-r_0)\xi]/J_0(R\xi)$. Putting $f(\xi)$ in
$\phi^{(s)}(r,z)$ and following \cite{R1}, evaluating the trivial $r_0$-integral and finally using some of the
properties of Bessel functions, we have the work function in the case of emission of electrons in the
transverse direction
\begin{equation}
W_f=-e^2\int_{\xi =0}^\infty \frac{\left (1+\frac{\lambda^2}{\xi^2}\right)^{1/2}- \frac{J_1(\xi
R)}{J_0(\xi R)}} {1-\left(1+\frac{\lambda^2}{\xi^2}\right)^{1/2}}d\xi
\end{equation}
This is a diverging integral for both $R\longrightarrow \infty$, or $R\longrightarrow 0$, i.e., the
value of the work function is independent of the transverse dimension of the cylindrical cell. For
$R\longrightarrow 0$, the second term on the numerator vanishes, whereas the rest can be integrated
analytically and found to be diverging in the upper limit. On the other hand, for $R\longrightarrow
\infty$, the second term becomes $\sim \tan(\xi R-\pi/4)$ (asymptotically). Which itself also diverges
for certain fixed values of the argument, but dose not cancel with the infinity from the other part of this
integral. Therefore in this limit also the work function becomes infinity. Although we have considered
here the extreme case of ultra-strong magnetic field, we do expect that even for low and moderate field
values, the work function corresponding to the electron emission in the transverse direction will be several 
orders of magnitudes larger than the corresponding longitudinal values. 
In fig.(2) the monotonically increasing straight line curve indicates the variation of $W_f$ with $B$. 
This graph clearly shows that for low field values, $W_f$ is a few ev in magnitude, which is of the same
orders of magnitudes with the experimentally known values for laboratory metals. 
However, in this model calculation, since the magnetic field is
assumed to be extremely high, we are unable to extend our calculation for very low field values. Also,
in this model we can not show the variation of $W_f$ from one kind of metal to another. 

As an application of this model we now obtain the density of ultra-relativistic electrons near the poles 
populated through field emission process. It was first introduced by Fowler and Nordheim \cite{R12} to
explain cold cathode emission phenomenon. The underline process is the tunneling of metallic electrons
through surface potential barrier. 
Following Fowler and Nordheim we have the electron number density
\begin{equation}
n_f=\frac{\alpha}{2\pi}\frac{E_F^{1/2}E_0 ^2}{(W_f+E_F)W_f ^{1/2}}\exp\left [- \frac{8\pi 
(2m_e)^{1/2}W_f ^{3/2}}{3hcE_0}\right ]
\end{equation}
where $\alpha$ is the fine structure constant, $E_F$ is the electron Fermi energy and $E_0$ is the field 
at the poles causing electron field emission. In the numerical evaluation of field electron density, 
we have taken both a constant $E_0$
($10^{12}$V/m), as given in the literature \cite{R2} and also as a function of magnetic field strength
at the poles, given in \cite{R5}. In fig.(3) we have shown the variation of
field electron density with the magnetic field strength. Monotonically increasing curves are for
magnetic field dependent $E_0$. The upper, middle and the lower curves are respectively for $n_e=10^{-5}$,
$10^{-4}$ and
$10^{-3}$ times normal nuclear density. Whereas the decreasing curves are for constant $E_0$,
with the same kind of variation for electron density. 

Next, we calculate the photo-electric current in the longitudinal direction. In the relativistic
case, in presence of strong quantizing magnetic field at zero temperature the photo-electric current $J$ is
given by 
\begin{equation}
J=\frac{eB}{2\pi^2}(E_F+E_\gamma -W_f)[1+\nu_{max}(\nu_{max}+1)]
\end{equation}
where $E_\gamma$ is the incident cosmic ray photon energy. In fig.(2) we have shown the variation of
photo-electric current density with the strength of magnetic field. For the sake of illustration, we
have taken $E_\gamma=1$GeV. The variation for $n_e=10^{-3}$, $10^{-4}$ and $10^{-5}$ times normal
nuclear density are shown by upper, middle and the lower monotonically decreasing curves respectively.
If the photon energy is increased, then it is obvious that $J$ will increase linearly with $E_\gamma$
for constant $B$ and $n_e$ (or $E_F$).

In conclusion, we mention that the main purpose of this letter is to show that (i) in presence of
strong quantizing magnetic field the work function becomes anisotropic, (ii) the transverse part is
infinitely large, (iii) the longitudinal part increases with the strength of magnetic field, (iv) low
field values of work function are more or less consistent with the tabulated values and finally, (v)
the use of longitudinal part of work function gives consistent result associated  with magneto sphere
electron density, reported in the published literature. 
In the present work, to the best of our knowledge, the anisotropic nature of work function in presence of
strong quantizing magnetic field is predicted for the first time. Further, the diverging character of
work function associated with the electron emission in the transverse direction is also obtained for
the first time, and so far our knowledge is concerned, it has not been reported in the published work.
We have also noticed that in the low magnetic field limit (within the limitation of this
model) the numerical values of work function are of the same order of magnitude with the known
laboratory data.
However, we are not able to compare our result with the variation from one metal to another. As an
application the use of magnetic field dependent work function to field emission gives results which
are consistent with the published theoretical results.

Since the results presented in this article are very much approximate in nature, we therefore conclude
with a beqautiful quotation from a very old but extremely interesting article by
Wigner and Bardeen \cite{wb}- "It is perhaps not quite 
superfluous to have, in addition to a more exact  calculations of a physical quantity, an approximate 
treatment which merely shows how the quantity in question is determined, and the lines along which a more 
exact calculation could be carried out. Such a treatment often leads to a simple formula by means of
which the magnitude of the quantity may be readily determined". 

Acknowledgement: We are thankful to Professor D. Schieber and Professor L. Schachter of Haifa for providing us
an online re-prient of \cite{R1}.

\begin{figure}
\psfig{figure=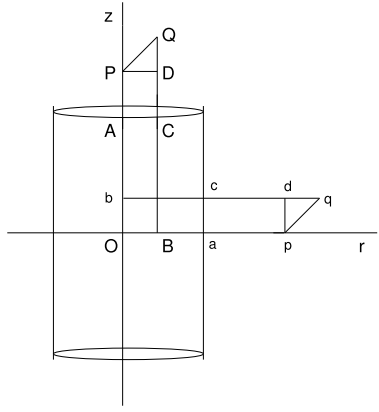,height=0.3\linewidth}
\caption{Schamatic diagram for electron emission along and orthogonal to the direction of magnetic
field.}
\end{figure}
\begin{figure}
\psfig{figure=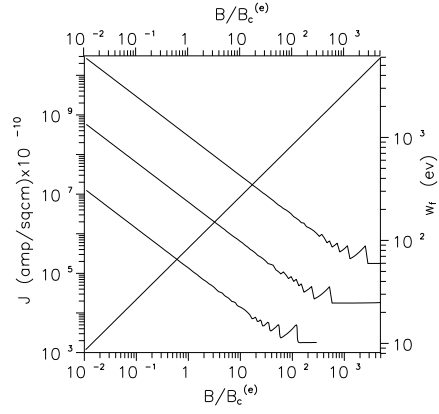,height=0.3\linewidth}
\caption{Variation of work function ($W_f$) with the magnetic field strength shown by monotonically
increasing straight line
curve. Variation of photo-electric current with magnetic field. Upper, middle and lower curves are
respectively for $10^{-3}$, $10^{-4}$ and $10^{-5}$ times normal nuclear density.}
\end{figure}
\begin{figure}
\psfig{figure=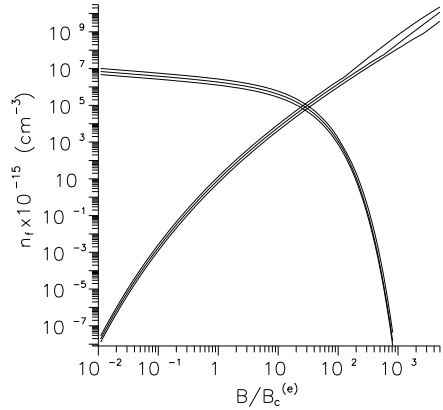,height=0.3\linewidth}
\caption{Variation of field electron density with magnetic field. Monotonically increasing curves are
for magnetic field dependent electric field, whereas the set of decreasing curves are for the constant
value of electric field as discussed in the text. For both the cases lower, middle and upper curves are
respectively for $10^{-3}$, $10^{-4}$ and $10^{-5}$ times normal nuclear density.}
\end{figure}
\end{document}